\documentclass[aps,prb,twocolumn,superscriptaddress,showpacs]{revtex4}

\usepackage{graphicx, bm, amsmath}

\newcommand{\me}{\mathrm{e}}
\newcommand{\mi}{\mathrm{i}}

\begin{document}

%\preprint{}

\title{Field-Induced Magnetic and Structural Domain Alignment in PrO$_2$}



\author{C.\,H. Gardiner}
\email{carol.webster@npl.co.uk}
\affiliation{National Physical Laboratory, Queens Road, Teddington, Middlesex, TW11 0LW, UK}

\author{A.\,T. Boothroyd}
\affiliation{Clarendon Laboratory, University of Oxford, Parks Road, Oxford, OX1 3PU, UK}

\author{M.\,J. McKelvy}
\affiliation{Center for Solid State Science, Arizona State University, Tempe, Arizona 85287-1704}

\author{G.\,J. McIntyre}
\affiliation{Institut Laue-Langevin, Bo{\^i}te Postale 156, F-38042 Grenoble C{\'e}dex 9, France}

\author{K. Proke\v s}
\affiliation{Hahn-Meitner Institut, SF-2, Glienicker Stra\ss e 100, D-14109 Berlin, Germany}

%\thanks{}


\date{\today}

\begin{abstract}
We present a neutron diffraction study of the magnetic structure
of single crystal PrO$_2$ under applied fields of 0--6\,T. As the
field is increased, changes are observed in the magnetic Bragg
intensities.  These changes are found to be irreversible when the
field is reduced, but the original intensities can be recovered by
heating to $T > 122$\,K, then re-cooling in zero field.  The
antiferromagnetic ordering temperature $T_{\rm N} = 13.5$\,K and
the magnetic periodicity are unaffected by the applied field.   We
also report measurements of the magnetic susceptibility of single
crystal PrO$_2$ under applied fields of 0--7\,T.  These show
strong anisotropy, as well as an anomaly at $T = 122 \pm 2$\,K
which coincides with the temperature $T_{\rm D} = 120 \pm 2$\,K at
which a structural distortion occurs.  For fields applied along
the [100] direction the susceptibility increases irreversibly with
field in the temperature range $T_{\rm N} < T < T_{\rm D}$.
However, for fields along [110] the susceptibility is independent
of field in this range.  We propose structural domain alignment,
which strongly influences the formation of magnetic domains below
$T_{\rm N}$, as the mechanism behind these changes.
\end{abstract}

%61.12.Ld Neutron diffraction.
%75.25.+z Spin arrangements in magnetically ordered materials (including neutron and spin-polarized electron studies, synchrotron-source x-ray scattering, etc.).
%75.30.Cr Saturation moments and magnetic susceptibilities.
%75.30.Kz Magnetic phase boundaries (including magnetic transitions, metamagnetism, etc.).

%\keywords{}

\maketitle


\section{Introduction}

In recent years there has been strong interest in Jahn-Teller and
orbital phenomena in compounds containing localized 4$f$ and 5$f$
electrons.  Among the simplest of these are the fluorite-structure
actinide dioxides UO$_2$ and NpO$_2$, which display complex
ordered phases at low temperatures involving coupled electric and
magnetic multipoles as well as (in the case of UO$_2$) a lattice
distortion. \cite{Santini, Paixao}

Unusual magnetic effects have also been observed in the lanthanide
dioxide PrO$_2$, which is isostructural with UO$_2$ and NpO$_2$ at
room temperature and exhibits antiferromagnetic ordering below
$T_{\rm N}$ = 13.5\,K.  Some years ago, PrO$_2$ was found to have
an anomalously small ordered moment in the antiferromagnetic
phase. \cite{Kern:1984} More recently we discovered a broad
continuum in the magnetic excitation spectrum probed by neutron
inelastic scattering, which we ascribed to Jahn-Teller
fluctuations involving the orbitally degenerate 4$f$ ground state
and dynamic distortions of the lattice. \cite{Boothroyd:2001} In a
separate neutron diffraction experiment \cite{Gardiner:2002:PrO2}
we found evidence that the antiferromagnetic structure contains a
component with twice the periodicity of the accepted magnetic
structure, and we have recently reported further neutron
diffraction studies \cite{Gardiner:PrO2Distortion} which reveal an
internal distortion of the fluorite structure below $T_{\rm D} =
120 \pm 2$\,K and a related distortion of the antiferromagnetic
structure below $T_{\rm N}$. These distortions result in a
doubling of both the crystallographic and magnetic unit cells
along one crystal axis. The magnetic structure is found to consist
of two components: a primary component whose unit cell is the same
as that of the undistorted crystal structure (referred to
hereafter as the ``type-I component"), and a secondary component
with a smaller ordered moment, whose unit cell is the same as that
of the distorted structure (referred to hereafter as the ``doubled
component").

In this paper we present neutron diffraction studies of the
magnetic structure and measurements of the magnetic susceptibility
of PrO$_2$ under applied magnetic fields of 0--7\,T. The neutron
diffraction experiments were designed to influence the alignment
of symmetry-equivalent magnetic domains, in order to determine
whether the type-I component of the magnetic structure was of the
multi-${\bf q}$ type.  These experiments produced unexpected
results which suggest that the alignment of the magnetic domains
is strongly influenced by a field-induced alignment of structural
domains in the distorted phase below $T_{\rm D}$.

\section{Magnetic structure under applied field}

\begin{figure}[!ht]
\begin{center}
\includegraphics{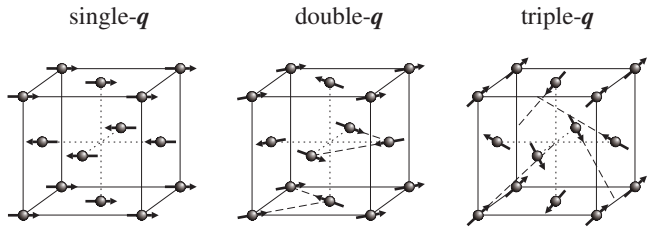}
\caption{Transverse multi-${\bf q}$ magnetic structures for
PrO$_2$.  The spheres are the Pr ions (the O ions are not shown in
this diagram).}
\label{fig:multi-q}
\end{center}
\end{figure}

As mentioned in the introduction, the magnetic structure of
PrO$_2$ contains two components, the dominant one being the type-I
antiferromagnetic structure which has a magnetic unit cell of the
same dimensions as the unit cell of the undistorted fluorite
crystallographic structure.  The secondary (doubled) component of
the magnetic structure is described in detail elsewhere.
\cite{Gardiner:PrO2Distortion} There are three possible type-I
spin arrangements consistent with existing neutron diffraction
data. These are the transverse multi-${\bf q}$ structures shown in
Fig. \ref{fig:multi-q}.  Longitudinal structures are ruled out by
comparison of their magnetic structure factors with the measured
Bragg intensities.  Neutron diffraction measurements in zero field
cannot distinguish between these three structures because they
give rise to identical magnetic structure factors when averaged
over symmetry-equivalent magnetic domains. One of the aims of the
experiments described here was to influence the populations of the
domains by cooling through $T_{\rm N}$ in a magnetic field, and
hence to distinguish between the three structures.

\subsection{Experimental Details}

Two neutron diffraction experiments were performed using the same
crystal in different orientations with respect to the magnetic
field.  The first experiment was performed on the E4 double-axis
single crystal diffractometer at the Berlin Neutron Scattering
Centre at the Hahn-Meitner Institute.  A flat pyrolytic graphite
(002) monochromator was used in combination with a graphite
filter. 40$^{\prime}$ collimators were placed before and after the
monochromator, but there was no collimation between the sample and
the detector.  A square aperture of 10\,mm $\times$ 10\,mm was
placed before the detector.  The incident neutron wavelength was
2.44\,\AA.

The second experiment was performed on the D10 single-crystal
diffractometer at the Institut Laue-Langevin.  This was operated
in double-axis mode with a position-sensitive detector. A
vertically curved pyrolytic graphite (002) monochromator was used
in combination with a graphite filter.  No collimators were used,
but a circular aperture of diameter 6\,mm was placed in the
incident beam before the sample, and a square aperture of 12\,mm
$\times$ 12\,mm was placed before the detector. The incident
neutron wavelength was 2.356(4)\,\AA.

For both experiments we used a single crystal sample of PrO$_2$
which was prepared some time ago by a hydrothermal procedure
\cite{McKelvy} and had a mass of $\approx$ 1\,mg.  For the first
experiment the crystal was mounted inside a vertical-field
superconducting cryomagnet with a temperature range of 2--300\,K
and a magnetic field range of 0--5\,T.  It was glued onto a thin
aluminium pin such that the [0$\bar{1}$1] direction was vertical
and hence parallel to the applied field. For the second experiment
the crystal was mounted inside a similar cryomagnet with a
slightly larger field range of 0--6\,T. It remained mounted on the
aluminium pin, but was re-aligned using an attachment that held it
at 45$^{\circ}$ so that the [001] direction was vertical.

\subsection{Measurements with ${\bf H} \parallel [0\bar{1}1]$}

We first describe measurements performed on the E4 diffractometer.
Five magnetic reflections of the type-I component of the magnetic
structure were accessible in the horizontal scattering plane of
the crystal ([0$\bar{1}$1] vertical). These were (100), (011),
(211), (122) and (300).  At the start of the experiment the
crystal was cooled to a temperature $T = 1.55$\,K with zero
applied field, and these five magnetic reflections plus several
structural Bragg reflections were measured by crystal rotation
($\omega$-scan).

\begin{figure}[!bt]
\begin{center}
\includegraphics{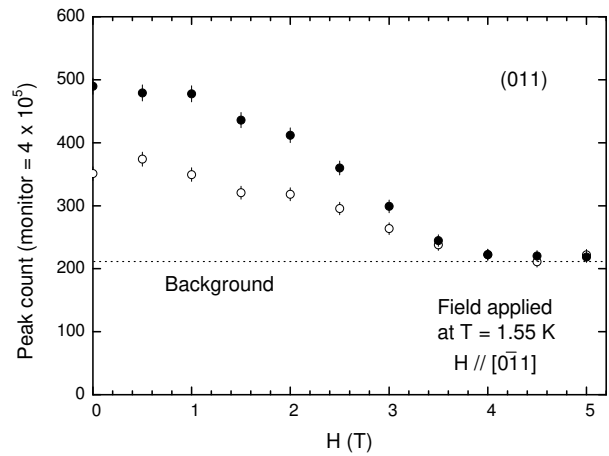}
\caption{Field dependence of the (011) magnetic Bragg reflection
before (open circles) and after (closed circles) the irreversible
increase in zero-field intensity.  The field was applied parallel
to [0$\bar{1}$1] and the temperature was held constant at $T =
1.55$\,K during both field sweeps.}
\label{fig:E4(011)vsHupanddown}
\end{center}
\end{figure}

After this we slowly increased the field from $H$ = 0\,T to $H$ =
5\,T in 0.5\,T steps (while remaining at $T$ = 1.55\,K) and
monitored the count at peak centre of the (011) reflection.  The
intensity decreased continuously with field, reducing almost to
zero at $H = 5$\,T.  As we did not warm above $T_{\rm N}$ during
the field sweep the disappearance of the (011) reflection
suggested a spin reorientation transition rather than a change in
the populations of the magnetic domains. To check that we could
regain the original intensity, we immediately warmed the sample to
$T = 20$\,K under $H = 5$\,T, removed the field, cooled the sample
back to $T = 1.55$\,K in zero field, and re-measured the five
type-I magnetic Bragg peaks. Surprisingly, the original
intensities were not recovered. $\omega$-scans revealed that the
intensities of all five peaks had increased by $\sim 100$\%
relative to the initial zero-field measurement. The intensities of
the (200) and (022) structural Bragg peaks were checked and found
to be the same as they had been before application of the field.
We then repeated the field scan from $H = 0$\,T to $H = 5$\,T at
$T = 1.55$\,K while counting at the centre of the (011) peak.  The
intensity was again found to reduce to zero (see Fig.\
\ref{fig:E4(011)vsHupanddown}).  By monitoring the zero-field
intensity of the (100) peak as a function of temperature, we
determined that the N\'eel temperature had not changed, remaining
at $T_{\rm N} = 13.5$\,K.

\begin{figure}[!ht]
\begin{center}
\includegraphics{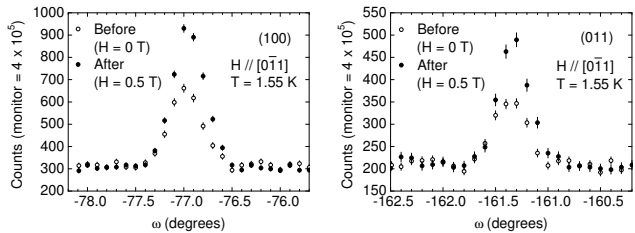}
\caption{$\omega$-scans of the (100) and (011) magnetic Bragg
reflections at $T$ = 1.55\,K, before (open circles) and after
(closed circles) application and removal, above $T_{\rm N}$, of a
5\,T magnetic field parallel to the [0$\bar{1}$1] direction.
Detector counts are normalised to a fixed incident beam monitor
count of $4 \times 10^5$, corresponding to a counting time of
$\sim 4$ minutes.} \label{fig:E4(100)and(011)peaks}
\end{center}
\end{figure}

To illustrate the increase in intensities of the five type-I
magnetic Bragg peaks we compare in Fig.\
\ref{fig:E4(100)and(011)peaks} the (100) and (011) peaks, measured
in approximately zero field, before and after application and
removal of the 5\,T field associated with the above temperature
cycling. We should mention that when the peaks were remeasured
after the irreversible increase in intensities the magnetic field
was actually set to $H = 0.5$\,T, rather than $H = 0$\,T, but
subsequent field scans showed almost no difference between 0\,T
and 0.5\,T (see Fig.\ \ref{fig:E4(011)vsHupanddown}).

In an attempt to recover the original intensities of the five
peaks the sample was heated to $T = 35$\,K and cooled back down to
1.55\,K in zero field.  However, the original intensities were not
recovered.  The sample was then heated to $T = 123$\,K and cooled
back down in zero field.  This time the original intensities were
recovered.

As mentioned in the introduction, PrO$_2$ undergoes a structural
distortion at $T_{\rm D}$ = 120\,$\pm$\,2\,K, which is accompanied
by an anomaly in the magnetic susceptibility.
\cite{Gardiner:PrO2Distortion} The fact that we were able to
recover the original intensities of the magnetic Bragg reflections
by heating to $T > T_{\rm D}$ suggests that the PrO$_2$ ground
state is metastable in the distorted phase.  From now on we will
refer to the Bragg intensities obtained by cooling through $T_{\rm
D}$ in zero field as the ``original intensities".

\begin{figure}[!ht]
\begin{center}
\includegraphics{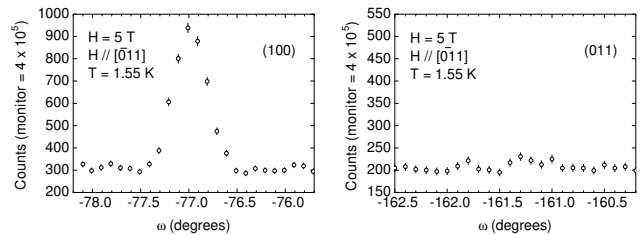}
\caption{$\omega$-scans of the (100) and (011) magnetic Bragg
reflections (normalised to a fixed incident beam monitor count),
measured at $H$ = 5\,T after cooling through $T_{\rm N}$ in the
field (${\bf H}\parallel$ {[0$\bar{1}$1]}).}
\label{fig:E4(100)and(011)5T}
\end{center}
\end{figure}

The intensities of three of the reflections, (100), (011) and
(211), were also measured by $\omega$-scan at $H = 5$\,T (after
cooling through $T_{\rm N}$ in the 5\,T field).  The (100) and
(211) reflections were found to have increased by $\sim 100$\%
from their original intensities at $H = 0$\,T, whereas the (011)
reflection was found to have decreased almost to the level of the
background as reported above (see Fig.\
\ref{fig:E4(011)vsHupanddown}). The (100) and (011) peaks are
shown in Fig.\ \ref{fig:E4(100)and(011)5T} for comparison with
those shown in Fig.\ \ref{fig:E4(100)and(011)peaks}.

Since the (100) peak did not disappear at $H = 5$\,T, we decided
to check its field dependence.  Starting from the original
intensity, we found that when the field was applied at $T =
1.55$\,K there was little change in intensity up to $H = 5$\,T.
However, when the crystal was warmed above $T_{\rm N}$ and cooled
to 1.55\,K each time the field was incremented we observed a
smooth increase in intensity with field. Furthermore, the increase
did not become irreversible until the applied field was larger
than $H = 3.5$\,T. Once the 100\% increase had been achieved, the
intensity of the (100) reflection was unaffected by the applied
field, whether the crystal was cooled through $T_{\rm N}$ in the
field or not.  The same was found to be true of the (211) and
(300) reflections. However, the (122) reflection behaved more like
the (011) reflection, decreasing to zero intensity at $H=5$\,T
when cooled through $T_{\rm N}$ in the applied field.

The main findings from the measurements described so far with
${\bf H} \parallel [0\bar{1}1]$ are (i) the disappearance of
certain magnetic Bragg reflections on application of a field $H
\sim 5$\,T, (ii) an irreversible 100\% increase in intensity of
all five peaks when the field is removed above $T_{\rm N}$ and the
sample re-cooled to 1.55\,K, (iii) recovery of the original
intensities after heating above the structural distortion
temperature $T_{\rm D} = 120$\,K and re-cooling in zero field.  We
also note that the irreversible increase in intensities can be
generated by applying a field $H \sim 5$\,T at 20\,K, then cooling
through $T_{\rm N}$ either before or after removing the field. The
application of fields up to 5\,T at $T = 1.55$\,K does not cause
the zero-field intensities to increase.

\subsection{Measurements with ${\bf H} \parallel [001]$}

The experiment performed on the D10 diffractometer was a repeat of
the study performed on E4, but this time with the field applied
parallel to the [001] direction.  Since the field was constrained
to be vertical in both experiments the crystal orientation had to
be changed for the D10 experiment.  This altered the scattering
plane and hence the range of accessible magnetic reflections. The
type-I reflections accessible with our chosen neutron wavelength
were (010), (110), (120), (030), (130) and (230).  A number of
half-integer reflections corresponding to the doubled component of
the magnetic structure were also accessible, but only the
$\left(\frac{1}{2} 1 0\right)$ and $\left(\frac{3}{2} 1 0\right)$
reflections had sufficient intensity to be measurable.

\begin{figure}[!ht]
\begin{center}
\includegraphics{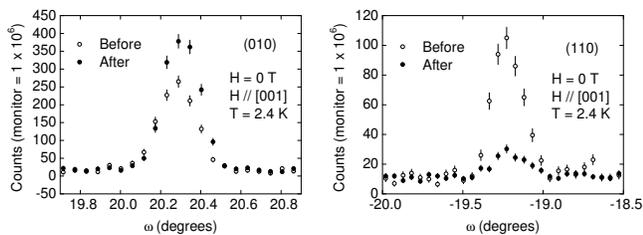}
\caption{The intensities of the (010) and (110) magnetic Bragg
reflections at $T$ = 2.4\,K, measured at $H$ = 0\,T before (open
circles) and after (closed circles) application and removal, above
$T_{\rm N}$, of a 5.7\,T magnetic field parallel to the [001]
direction. A monitor count of $1 \times 10^6$ corresponds to a
counting time of $\sim 2$ minutes.}
\label{fig:D10(010)and(110)peaks}
\end{center}
\end{figure}

We measured these eight peaks by $\omega$-scan at $T = 2.4$\,K in
zero applied field.  We then warmed to 18\,K, applied a field of
5.7\,T, and re-cooled through $T_{\rm N}$ in the applied field.
Before re-measuring the peaks we warmed to 18\,K and removed the
field before cooling back to $T$ = 2.4\,K (this second
warming/cooling cycle was performed to reproduce the conditions
under which the irreversible increase in intensities was first
observed in the E4 experiment, i.e. field removed above $T_{\rm
N}$).  The repeat $\omega$-scans revealed that the (110) and (130)
peaks reduced by $\sim 80$\%, while the other type-I peaks
increased by $\sim 40$\%. The $\left(\frac{1}{2} 1 0\right)$ and
$\left(\frac{3}{2} 1 0\right)$ peaks increased by $\sim 20$\% and
30\% respectively. Figure \ref{fig:D10(010)and(110)peaks} shows
the intensities of the (010) and (110) peaks before and after
application and removal of the 5.7\,T field associated with the
above temperature cycling.

We then measured the (010), (110) and (230) peaks at $T$ = 2.4\,K,
$H \ge 5$\,T after cooling through $T_{\rm N}$ in the field. Their
intensities remained approximately the same as those observed at
zero field after application and removal of the 5.7\,T field.

\begin{figure}[!ht]
\begin{center}
\includegraphics{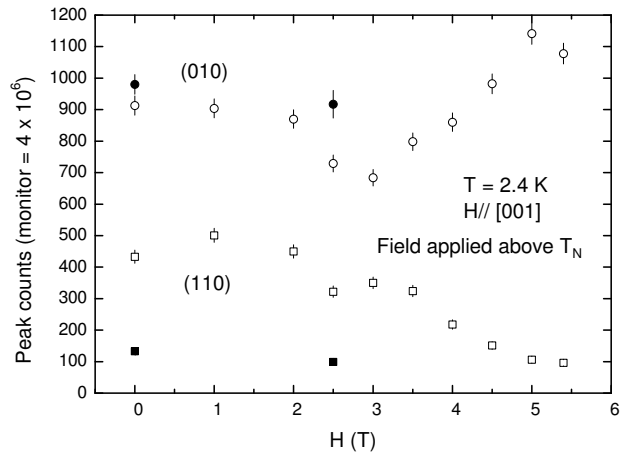}
\caption{Field dependence of the (010) and (110) magnetic Bragg
reflections at peak centre before (open symbols) and after (closed
symbols) the irreversible change in peak intensities. The magnetic
field was applied parallel to the [001] direction, and the crystal
was warmed above $T_{\rm N}$ and cooled back to $T = 2.4$\,K each
time the field was changed. Note that the monitor count for this
data is four times that of the data displayed in Figure
\ref{fig:D10(010)and(110)peaks}.}
\label{fig:D10(010)and(110)vsH}
\end{center}
\end{figure}

Figure \ref{fig:D10(010)and(110)vsH} shows a plot of the count at
peak centre for the (010) and (110) reflections as a function of
applied field before and after the change in intensities.  Each
time the field was changed the crystal was warmed above $T_{\rm
N}$ and cooled back to $T = 2.4$\,K.  Once the maximum field of
5.7\,T had been applied, further changes in the applied field had
little effect on the peak intensities.

In agreement with expectations following the E4 experiment, we
found that application and removal of a large magnetic field
(5.7\,T) above $T_{\rm N}$, followed by zero-field cooling caused
an irreversible change in the intensities of the magnetic Bragg
peaks.  We also found that the original intensities could only be
regained by heating the crystal to $T > T_{\rm D}$ and re-cooling
through $T_{\rm D}$ in zero field.  The intensities of the (020)
and (220) structural Bragg reflections were checked at $T$ =
2.4\,K, $H$ = 0\,T before and after application of the 5.7\,T
field. Again, these did not change. There was no change in the
N\'eel temperature either, this remaining at $T_{\rm N} = 13.4$\,K
for both integer and half-integer magnetic reflections.

The most striking result from the measurements described above is
that the changes in the peak intensities are different from those
observed during the E4 experiment.  In the E4 experiment (when $H$
was applied along the [0$\bar{1}$1] direction) all the peak
intensities rose by $\sim 100$\% after application and removal of
a 5\,T field, whereas in the D10 experiment (when $H$ was applied
along the [001] direction) the relative peak intensities changed:
some peaks increased by $\sim 40$\% while others decreased by
$\sim 80$\%.

\subsection{Data analysis}

We obtained the integrated intensity of each magnetic Bragg
reflection measured in the experiments described above by fitting
a Gaussian profile to each peak, calculating its area and
correcting for the geometrical Lorentz factor $L = \sin 2\theta$,
where $\theta$ is the Bragg angle.  Tables \ref{tab:E4intensities}
and \ref{tab:D10intensities} list the integrated, corrected
intensities and compare them with the square of the expected
magnetic structure factor $\langle|F_{\rm M}({\bf Q})|^2\rangle$
for a type-I antiferromagnetic structure under ambient conditions,
i.e. where all the magnetic domains are equally populated.  The
magnetic structure factors were calculated using the formula

\begin{equation}
\langle|F_{\rm M}({\bf Q})|^2\rangle =
\sum_{\alpha\beta}\left\langle\left(\delta_{\alpha\beta} -
\hat{Q}_{\alpha}\hat{Q}_{\beta}\right)F_{\rm M}^{\alpha}({\bf
Q})F_{\rm M}^{\beta}({\bf Q})\right\rangle, \label{eq:average}
\end{equation}

\noindent where the summation indices $\alpha$ and $\beta$ run
over the cartesian co-ordinates $x,y$ and $z$,
$\delta_{\alpha\beta}$ is the Kronecker delta, $\hat{Q}_{\alpha}$
is the $\alpha$-component of the unit scattering vector and
$\langle \rangle$ denotes an average over all symmetry-equivalent
magnetic domains.  $F_{\rm M}^{\alpha}({\bf Q})$ is given by

\begin{equation}
F_{\rm M}^{\alpha}({\bf Q}) = f({\bf
Q})\sum_{j}\hat\mu_j^{\alpha}\me^{\mi{\bf Q}.{\bf
r}_j}\me^{-W_j(Q,T)},
\end{equation}

\noindent where the summation index $j$ runs over all the magnetic
atoms in the magnetic unit cell, $\hat\mu_j^{\alpha}$ is the
$\alpha$-component of a unit vector in the direction of the
magnetic moment of the $j$th magnetic atom, ${\bf r}_j$ is the
position of the $j$th magnetic atom within the magnetic unit cell,
$f({\bf Q})$ is the magnetic form factor of the Pr$^{4+}$ ion and
$\me^{-W_j(Q,T)}$ is the Debye-Waller factor (we set this equal to
1, since the measurements were made at low temperatures).

The integrated intensities of the magnetic Bragg reflections are
proportional to $\langle|F_{\rm M}({\bf Q})|^2\rangle$ and the
volume of the sample.  For ease of comparison the intensities of
the observed magnetic reflections listed in tables
\ref{tab:E4intensities} and \ref{tab:D10intensities} have been
normalised to the original intensity of the (100) reflection.
Similarly, the $\langle|F_{\rm M}({\bf Q})|^2\rangle$ have been
normalised to the $\langle|F_{\rm M}({\bf Q})|^2\rangle$ of the
(100) reflection.

\begin{table}[!ht]
\renewcommand{\arraystretch}{1.5}
\begin{center}
\begin{tabular}{|c||c|c|c||c|}
\hline
\rule[1mm]{0mm}{11pt} & \multicolumn{3}{c||}{${\bf H}\parallel$ [0$\bar{1}$1]} & \raisebox{-6pt}{Domain-} \\
\cline{2-4}
\rule[1mm]{0mm}{11pt}  Reflection & Before    & \multicolumn{2}{c||}{After} & \raisebox{3pt}{averaged} \\
\cline{2-4}
\rule[1mm]{0mm}{11pt}       & $H=0$      & $H=0.5$\,T & $H=5$\,T   & \raisebox{6pt}{$\left\langle|F_{\rm M}({\bf Q})|^2\right\rangle$} \\
\hline\hline
\rule[1mm]{0mm}{12pt} (100) & 1.00 $\pm$ 0.05 & 1.87 $\pm$ 0.05 & 1.93 $\pm$ 0.06 & 1.00 \\
\hline
\rule[1mm]{0mm}{12pt} (011) & 0.56 $\pm$ 0.04 & 1.00 $\pm$ 0.06 & 0.09 $\pm$ 0.05 & 0.48 \\
\hline
\rule[1mm]{0mm}{12pt} (211) & 0.63 $\pm$ 0.07 & 1.34 $\pm$ 0.09 & 1.18 $\pm$ 0.09 & 0.65 \\
\hline
\rule[1mm]{0mm}{12pt} (122) & 0.47 $\pm$ 0.07 & 0.91 $\pm$ 0.08 & *0.51 $\pm$ 0.26 & 0.38 \\
\hline
\rule[1mm]{0mm}{12pt} (300) & 0.56 $\pm$ 0.07 & 1.17 $\pm$ 0.09 & *1.67 $\pm$ 0.31 & 0.69 \\
\hline
\end{tabular}
\caption{Comparison between integrated intensities of magnetic
peaks (corrected for the Lorentz factor) before and after
application of a 5\,T magnetic field along [0$\bar{1}$1]. The
intensities have been normalised to the original intensity of the
(100) reflection. The $\langle|F_{\rm M}({\bf Q})|^2\rangle$ have
been normalised to the $\langle|F_{\rm M}({\bf Q})|^2\rangle$ of
the (100) reflection. The values marked with a * have been
calculated from measurements of the count at peak centre, since no
$\omega$-scans were made at these positions.}
\label{tab:E4intensities}
\end{center}
\end{table}

\begin{table}[!ht]
\renewcommand{\arraystretch}{1.5}
\begin{center}
\begin{tabular}{|c||c|c|c||c|}
\hline
\rule[1mm]{0mm}{11pt} & \multicolumn{3}{c||}{${\bf H}\parallel$ [001]}        &   \raisebox{-6pt}{Domain-}     \\
\cline{2-4}
\rule[1mm]{0mm}{11pt}  Reflection  & Before  & \multicolumn{2}{c||}{After} & \raisebox{3pt}{averaged} \\
\cline{2-4}
\rule[1mm]{0mm}{11pt}                              & $H=0$       & $H=0$ & $H\approx 5$\,T & \raisebox{6pt}{$\left\langle|F_{\rm M}({\bf Q})|^2\right\rangle$} \\
\hline\hline
\rule[1mm]{0mm}{12pt} (010)                        & 1.00 $\pm$ 0.09 & 1.43 $\pm$ 0.10 & 1.38 $\pm$ 0.10 & 1.00 \\
\hline
\rule[1mm]{0mm}{12pt} (110)                        & 0.50 $\pm$ 0.06 & 0.10 $\pm$ 0.03 & 0.09 $\pm$ 0.05 & 0.48 \\
\hline
\rule[1mm]{0mm}{12pt} (120)                        & 0.53 $\pm$ 0.11 & 0.76 $\pm$ 0.10 &                 & 0.49 \\
\hline
\rule[1mm]{0mm}{12pt} (030)                        & 0.57 $\pm$ 0.08 & 0.79 $\pm$ 0.08 &                 & 0.69 \\
\hline
\rule[1mm]{0mm}{12pt} (130)                        & 0.26 $\pm$ 0.06 & 0.07 $\pm$ 0.02 &                 & 0.33 \\
\hline
\rule[1mm]{0mm}{12pt} (230)                        & 0.47 $\pm$ 0.07 & 0.65 $\pm$ 0.08 & 0.76 $\pm$ 0.16 & 0.49 \\
\hline
\rule[1mm]{0mm}{12pt} $\left(\frac{1}{2}10\right)$ & 0.08 $\pm$ 0.02 & 0.10 $\pm$ 0.02 &                 &      \\
\hline
\rule[1mm]{0mm}{12pt} $\left(\frac{3}{2}10\right)$ & 0.10 $\pm$ 0.02 & 0.13 $\pm$ 0.03 &                 &      \\
\hline
\end{tabular}
\caption{Comparison between integrated intensities of magnetic
peaks (corrected for the Lorentz factor) before and after
application of a 5.7\,T magnetic field along [001]. The
intensities have been normalised to the original intensity of the
(010) reflection. The $\langle|F_{\rm M}({\bf Q})|^2\rangle$ have
been normalised to the $\langle|F_{\rm M}({\bf Q})|^2\rangle$ of
the (010) reflection.}
\label{tab:D10intensities}
\end{center}
\end{table}

For ${\bf H}\parallel$ [0$\bar{1}$1] the relative intensities of
the magnetic Bragg reflections agree with the domain-averaged
$\langle|F_{\rm M}({\bf Q})|^2\rangle$ both before and after the
irreversible increase in intensities. However, for ${\bf
H}\parallel$ [001] the relative peak intensities only agree with
the domain-averaged $\langle|F_{\rm M}({\bf Q})|^2\rangle$ before
the irreversible change.

\subsection{Summary}

To summarise the results of the neutron diffraction studies on
single crystal PrO$_2$, we have observed an irreversible change in
the magnetic Bragg intensities following the application and
removal of a large magnetic field.  The field required to induce
this change is between 3.5\,T and 5\,T when ${\bf H}\parallel$
[0$\bar{1}$1] and $\leq$ 5.7\,T when ${\bf H}\parallel$ [001]. To
recover the original intensities the crystal must be heated to $T
> T_{\rm D}$ and re-cooled through $T_{\rm D}$ in zero field.
The experimental results suggest that the field must be applied
above $T_{\rm N}$ to induce the irreversible change in
intensities.  When the field is applied parallel to the
[0$\bar{1}$1] direction, certain peaks increase by $\sim 100$\%
while others disappear.  After removal of the field, all peaks are
found to have increased by $\sim 100$\% relative to their original
intensities.  When the field is applied parallel to [001] some of
the type-I peaks increase by $\sim 40$\% while others decrease by
$\sim 80$\%.  The half-integer peaks increase by 20--30\%.  These
changes are preserved when the field is removed.  The intensities
of the structural Bragg reflections are unaffected by fields
applied along either the [0$\bar{1}$1] direction or the [001]
direction, and $T_{\rm N}$ remains unchanged.

\section{Influence of magnetic field on domain populations}

\label{sec:domains}

We now attempt to interpret the observations summarised above by
considering the possible influence of an applied magnetic field on
the populations of structural and magnetic domains.

One of the original aims of the neutron diffraction experiments
described above was to influence the populations of magnetic
domains by cooling through $T_{\rm N}$ in a magnetic field.  By
applying the field along certain symmetry directions we hoped to
break the symmetry of the crystal in such a way that certain
domains would be energetically favoured.  In the single-${\bf q}$
and double-${\bf q}$ magnetic structures some domains have zero
magnetic structure factors.  Therefore, altering the populations
of the domains can alter the magnetic Bragg intensities.  However,
in the triple-${\bf q}$ structure all domains have identical
magnetic structure factors, so in this case the Bragg intensities
will be unaltered by changes in the domain populations.

The fact that we have observed changes in the intensities of the
magnetic Bragg peaks after application of a magnetic field along
two different symmetry directions rules out the triple-${\bf q}$
structure.  We can also rule out the double-${\bf q}$ structure by
calculating the single-domain magnetic structure factors for the
observed reflections.  We find no single domain whose squared
magnetic structure factor is more than 50\% larger than its value
when averaged over all symmetry-equivalent domains.  Thus the
double-${\bf q}$ structure cannot account for the 100\% increase
in intensities observed when the field is applied along
[0$\bar{1}$1].  However, to prove that the structure is
single-${\bf q}$ and that the increase in intensities observed is
due to preferential population of certain domains we must
determine whether the increased intensities are consistent with
the magnetic structure factors of the domains we expect to be
favoured.

The single-${\bf q}$ transverse\footnote{The longitudinal
multi-${\bf q}$ structures are ruled out because they predict
$\langle|F_{\rm M}({\bf Q})|^2\rangle$ = 0 for the (100)
reflection, which is not observed either before or after
application of a large magnetic field.} structure has six
symmetry-equivalent domains, as the ordering wavevector can be
along [100], [010] or [001], and the spins can point along either
of the two directions mutually perpendicular to the ordering
vector. Tables \ref{tab:singledomainstrucfactorsE4} and
\ref{tab:singledomainstrucfactorsD10} show the single-domain
$|F_{\rm M}({\bf Q})|^2$ calculated in these six domains for the
magnetic Bragg peaks observed in the E4 and D10 experiments
respectively. We use $D_{\alpha}^{\beta}$ to denote a domain with
ordering vector along the $\alpha$-direction and spins along the
$\beta$-direction.

\begin{table}[!ht]
\renewcommand{\arraystretch}{1.5}
\begin{center}
\begin{tabular}{|c||c|c|c|c|c|c||c|}
\hline
\rule[1mm]{0mm}{11pt} & \multicolumn{6}{c||}{\rule[1mm]{0mm}{11pt} Single-domain $|F_{\rm M}({\bf Q})|^2$} & \raisebox{-2pt}{Domain-averaged} \\
\cline{2-7}
\rule[1mm]{0mm}{11pt} Reflection & $D_a^b$ & $D_a^c$ & $D_b^a$ & $D_b^c$ & $D_c^a$ & $D_c^b$ & \rule[1mm]{0mm}{11pt} \raisebox{3pt}{$\left\langle|F_{\rm M}({\bf Q})|^2\right\rangle$} \\
\hline\hline
\rule[1mm]{0mm}{12pt} (100)      & 3.00    & 3.00    & 0.00    & 0.00    & 0.00    & 0.00    & 1.00 \\
\hline
\rule[1mm]{0mm}{12pt} (011)      & 1.42    & 1.42    & 0.00    & 0.00    & 0.00    & 0.00    & 0.48 \\
\hline
\rule[1mm]{0mm}{12pt} (211)      & 1.96    & 1.96    & 0.00    & 0.00    & 0.00    & 0.00    & 0.65 \\
\hline
\rule[1mm]{0mm}{12pt} (122)      & 1.14    & 1.14    & 0.00    & 0.00    & 0.00    & 0.00    & 0.38 \\
\hline
\rule[1mm]{0mm}{12pt} (300)      & 2.06    & 2.06    & 0.00    & 0.00    & 0.00    & 0.00    & 0.69 \\
\hline
\end{tabular}
\caption{Comparison between single-domain and domain-averaged
$|F_{\rm M}({\bf Q})|^2$ for magnetic Bragg reflections observed
in the E4 experiment (where the magnetic field was applied along
[0$\bar{1}$1]).  The single-domain $|F_{\rm M}({\bf Q})|^2$ have
been normalised to the domain-averaged $\langle|F_{\rm M}({\bf
Q})|^2\rangle$ of the (100) reflection. $D_{\alpha}^{\beta}$
denotes a domain with ordering vector along the $\alpha$-direction
and spins along the $\beta$-direction.}
\label{tab:singledomainstrucfactorsE4}
\end{center}
\end{table}

\begin{table}[!ht]
\renewcommand{\arraystretch}{1.5}
\begin{center}
\begin{tabular}{|c||c|c|c|c|c|c||c|}
\hline
\rule[1mm]{0mm}{11pt} & \multicolumn{6}{c||}{\rule[1mm]{0mm}{11pt} Single-domain $|F_{\rm M}({\bf Q})|^2$} & \raisebox{-2pt}{Domain-averaged} \\
\cline{2-7}
\rule[1mm]{0mm}{11pt} Reflection & $D_a^b$ & $D_a^c$ & $D_b^a$ & $D_b^c$ & $D_c^a$ & $D_c^b$ & \rule[1mm]{0mm}{11pt} \raisebox{3pt}{$\left\langle|F_{\rm M}({\bf Q})|^2\right\rangle$} \\
\hline\hline
\rule[1mm]{0mm}{12pt} (010)      & 0.00    & 0.00    & 3.00    & 3.00    & 0.00    & 0.00    & 1.00 \\
\hline
\rule[1mm]{0mm}{12pt} (110)      & 0.00    & 0.00    & 0.00    & 0.00    & 1.42    & 1.42    & 0.48 \\
\hline
\rule[1mm]{0mm}{12pt} (120)      & 0.49    & 2.46    & 0.00    & 0.00    & 0.00    & 0.00    & 0.49 \\
\hline
\rule[1mm]{0mm}{12pt} (030)      & 0.00    & 0.00    & 2.06    & 2.06    & 0.00    & 0.00    & 0.69 \\
\hline
\rule[1mm]{0mm}{12pt} (130)      & 0.00    & 0.00    & 0.00    & 0.00    & 1.77    & 0.20    & 0.33 \\
\hline
\rule[1mm]{0mm}{12pt} (230)      & 0.00    & 0.00    & 1.20    & 1.74    & 0.00    & 0.00    & 0.49 \\
\hline
\end{tabular}
\caption{Comparison between single-domain and domain-averaged
$|F_{\rm M}({\bf Q})|^2$ for magnetic Bragg reflections observed
in the D10 experiment (where the magnetic field was applied along
[001]).  The single-domain $|F_{\rm M}({\bf Q})|^2$ have been
normalised to the domain-averaged $\langle|F_{\rm M}({\bf
Q})|^2\rangle$ of the (010) reflection. $D_{\alpha}^{\beta}$
denotes a domain with ordering vector along the $\alpha$-direction
and spins along the $\beta$-direction.}
\label{tab:singledomainstrucfactorsD10}
\end{center}
\end{table}

If we assume that the exchange interactions between the individual
atomic spins are stronger than their interactions with the applied
field, then for an antiferromagnetic structure we expect the
domains with the lowest magnetic energy to be those whose spins
lie perpendicular to the applied field.  We therefore expect such
domains to be favoured over those whose spins have components
parallel and antiparallel to the field.  If we use this assumption
to predict which magnetic domains should be favoured in PrO$_2$,
we expect domains $D_b^a$ and $D_c^a$ to be favoured equally for
${\bf H} \parallel [0\bar{1}1]$.  However, the magnetic structure
factors in these domains are zero for all the peaks measured in
the E4 experiment, in contradiction to the observed irreversible
increase in intensities. Similarly, for ${\bf H}
\parallel [001]$ we would expect domains $D_a^b$, $D_b^a$, $D_c^a$ and $D_c^b$ to be favoured
equally. However, if we average the magnetic structure factors for
these domains we find that they do not agree with the observed
intensity changes.

The above observations suggest that our assumptions concerning the
dominant spin interactions are incorrect. By inspection of table
\ref{tab:singledomainstrucfactorsE4} we find better agreement
between the structure factors and increased intensities if domains
$D_a^b$ and $D_a^c$ are favoured for ${\bf H} \parallel
[0\bar{1}1]$. Rather than having \emph{spins} perpendicular to the
applied field, these domains have \emph{ordering vectors}
perpendicular to the applied field.  If these domains are
favoured, the intensities of all the observed peaks should
increase by a factor of 3.  The observed increase is a factor of
$\sim 2$, but it is possible that this is due to incomplete
depopulation of the disfavoured domains. Similarly, by inspection
of table \ref{tab:singledomainstrucfactorsD10} we find better
agreement between the structure factors and observed intensities
if domains $D_a^b$, $D_a^c$, $D_b^a$ and $D_b^c$ are favoured for
${\bf H} \parallel [001]$.  Again, these domains have ordering
vectors (rather than spins) perpendicular to the applied field. If
we average their squared magnetic structure factors we find that
the (110) and (130) peaks are expected to disappear, while the
rest are expected to increase by 50\%.  The observed decrease of
the (110) and (130) intensities by $\sim 80$\% and the increase of
the other peaks by $\sim 40$\% are consistent with a small portion
of the crystal remaining populated with the disfavoured domains.

In the D10 experiment we also observed the
$\left(\frac{1}{2}10\right)$ and $\left(\frac{3}{2}10\right)$
reflections to increase by 20--30\% after application and removal
of a large field parallel to [001].  These reflections arise from
the doubled component of the magnetic structure.
\cite{Gardiner:PrO2Distortion}  If we calculate the single-domain
magnetic structure factors of the $\left(\frac{1}{2}10\right)$ and
$\left(\frac{3}{2}10\right)$ reflections, we find that they are
non-zero only in domains where the unit cell is doubled along the
$a$-direction.  Since the field was applied along [001], i.e.\ the
$c$-direction in the D10 experiment, we suggest that the applied
field favours domains of the doubled component of the magnetic
structure whose unit cells are doubled along the $a$- and
$b$-directions, i.e.\ directions perpendicular to ${\bf H}$.  The
squared magnetic structure factors of the
$\left(\frac{1}{2}10\right)$ and $\left(\frac{3}{2}10\right)$
reflections averaged over these domains are 50\% larger than the
structure factors averaged over all domains, in rough agreement
with the observed increase in intensities of the peaks.

We have shown that magnetic domain alignment can account for the
observed changes in magnetic Bragg intensities following
application and removal of a large magnetic field. However, we
have not yet discussed why the changes are irreversible below
$T_{\rm D}$.  We propose that this is because certain domains of
the distorted crystal structure are energetically preferred in an
applied field. In other words, the application of a magnetic field
at constant temperature below $T_{\rm D}$ causes a change in the
structural domain population. Intuitively, it feels somewhat
implausible that a 5\,T magnetic field could cause a significant
repopulation of structural domains at low temperatures.  However,
we must remember that the structural distortion is purely
internal, so does not produce any macroscopic strain.  All that is
required is that the oxygen atoms move the small distance between
equivalent shifted sites within the unit cell.  Since it is likely
that the cooperative Jahn-Teller distortion at $T_{\rm D}$ is
accompanied by ordering of the Pr orbitals, we propose that the
mechanism by which the field influences the structural domains is
a coupling between the applied field and the Pr orbital magnetic
moments.  Our analysis indicates that the favoured structural
domains are those whose unit cells are doubled along directions
perpendicular to the applied field. The existence of a doubled
component of the magnetic structure indicates a strong coupling
between the lattice and the magnetic ordering, possibly via the Pr
spin-orbit interaction. We therefore propose that the formation of
magnetic domains on cooling through $T_{\rm N}$ is influenced
predominantly by the underlying alignment of the structural
domains.

From our analysis of the favoured type-I magnetic domains it
appears that coupling between the type-I component and the doubled
component of the magnetic structure causes the ordering vector of
the type-I component to lie parallel to the direction along which
the unit cell is doubled. The strong influence of the structural
domain alignment on the magnetic ordering of both the type-I
component and the doubled component would account for the
irreversibility of the changes in the magnetic Bragg intensities
below $T_{\rm D}$.

\begin{figure}[!ht]
\begin{center}
\includegraphics{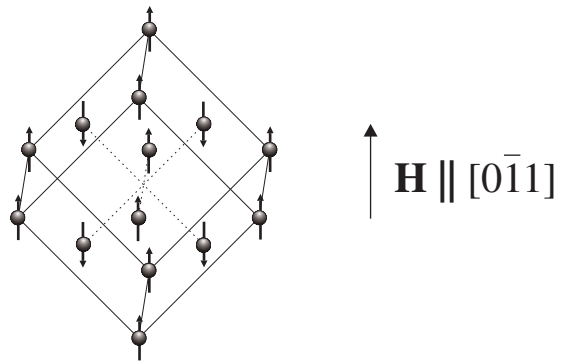}
\caption{Reorientated spin configuration of the type-I component
of the magnetic structure at $H = 5$\,T.}
\label{fig:spinreorientation}
\end{center}
\end{figure}

Finally we recall that in the E4 experiment the (011) and (122)
reflections disappeared under the influence of a 5\,T magnetic
field along [0$\bar{1}$1].  This occurred both when the field was
applied at constant $T = 1.55$\,K and when the crystal was cooled
through $T_{\rm N}$ in the field.  The reflections reappeared when
the field was removed.  To explain these observations we propose
that a reversible spin-reorientation transition occurs in the
type-I component between $H$ = 0\,T and $H$ = 5\,T for ${\bf H}
\parallel [0\bar{1}1]$. This causes the spins to rotate parallel
and antiparallel to the field as shown in Fig.\
\ref{fig:spinreorientation}.  We arrived at this spin
configuration by trying various possibilities until we found one
whose squared single-domain magnetic structure factors were in
agreement with the relative intensities of the magnetic peaks
observed at $H = 5$\,T (see Table \ref{tab:spinreorientation}).
The configuration appears counterintuitive because an
antiferromagnetic structure would normally minimise its energy by
rotating its spins perpendicular to the field. However, our
experimental results indicate a strong interaction between the
lattice and the magnetism which could alter this simple picture.

\begin{table}[!ht]
\renewcommand{\arraystretch}{1.5}
\begin{center}
\begin{tabular}{|c||c|c|c|c|c|c||c|}
\hline
\rule[1mm]{0mm}{11pt} Reflection & Intensity at $H$ = 5\,T & Single-domain $|F_{\rm M}({\bf Q})|^2$ \\
\hline\hline
\rule[1mm]{0mm}{12pt} (100)      & 1.93 $\pm$ 0.06    & 3.00 \\
\hline
\rule[1mm]{0mm}{12pt} (011)      & 0.09 $\pm$ 0.05    & 0.00 \\
\hline
\rule[1mm]{0mm}{12pt} (211)      & 1.18 $\pm$ 0.09    & 1.57 \\
\hline
\rule[1mm]{0mm}{12pt} (122)      & *0.51 $\pm$ 0.26   & 0.23 \\
\hline
\rule[1mm]{0mm}{12pt} (300)      & *1.67 $\pm$ 0.31   & 2.05 \\
\hline
\end{tabular}
\caption{Comparison between integrated intensities of magnetic
peaks (corrected for the Lorentz factor) measured under a field of
5\,T parallel to [0$\bar{1}$1] and single-domain $|F_{\rm M}({\bf
Q})|^2$ for the structure shown in Fig.\
\ref{fig:spinreorientation}. The intensities have been normalised
to the original intensity of the (100) reflection. The $|F_{\rm
M}({\bf Q})|^2$ have been normalised to the domain-averaged
$\langle|F_{\rm M}({\bf Q})|^2\rangle$ of the (100) reflection for
the type-I magnetic structure. The values marked with a
* have been calculated from measurements of the count at peak
centre, since no $\omega$-scans were made at these positions.}
\label{tab:spinreorientation}
\end{center}
\end{table}

\section{Susceptibility measurements}

We now present measurements of the magnetic susceptibility of
single crystal PrO$_2$.  These measurements were made using a
commercial SQUID magnetometer with a vertical field range of
0--7\,T and a temperature range of 2--300\,K.  The susceptibility
was measured as a function of temperature and field, with the
field applied along two different symmetry directions.  The aim
was to probe the bulk magnetisation of the crystal as a function
of magnetic field to observe the effects of domain alignment and
spin reorientation.

\subsection{Experimental Details}

We used two single crystals of PrO$_2$ for the susceptibility
measurements.  Both were taken from the same batch as the sample
used in the neutron diffraction experiments, and each had a mass
of $< 1$\,mg.  The crystals were mounted in plastic sample holders
with the [110] and [001] directions vertical (hence parallel to
the applied field).  Their orientations were estimated to be
accurate to within 5\%.  All measurements were made using the
reciprocating sample option (RSO) which causes the sample holder
to oscillate in the vertical direction.

\subsection{Measurements and results}

Here we present measurements of the magnetic susceptibility
performed with the applied field parallel to the [110] and [001]
directions. Results obtained for the two different field
directions are compared and discussed.

It should be noted that the susceptibility measurements presented
in this section are not normalized to the mass of the crystals
because the crystals were too small to be weighed accurately.
Hence, the susceptibility data taken in different field directions
should not be compared quantitatively.

\begin{figure}[!ht]
\begin{center}
\includegraphics{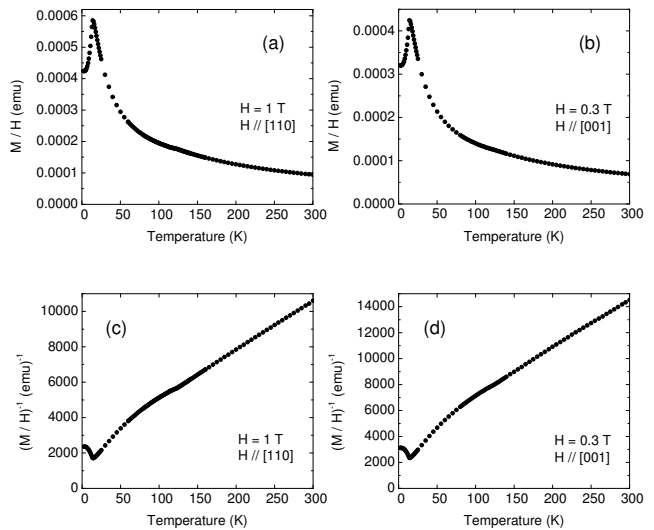}
\caption{Temperature dependence of the magnetic susceptiblity of
PrO$_2$ when the magnetic field is applied parallel to the [110]
and [001] directions.  Plots (c) and (d) show the inverse
susceptibilities derived from the data in (a) and (b)
respectively.}
\label{fig:(110)and(100)sus2-300K}
\end{center}
\end{figure}

Figures \ref{fig:(110)and(100)sus2-300K}(a) and
\ref{fig:(110)and(100)sus2-300K}(b) show the temperature
dependence of the magnetic susceptibility in the range $T$ =
2--300\,K when a field of $H = 1$\,T is applied parallel to the
[110] direction and when a field of $H = 0.3$\,T is applied
parallel to the [001] direction.  Figures
\ref{fig:(110)and(100)sus2-300K}(c) and
\ref{fig:(110)and(100)sus2-300K}(d) show the same data as in
Figs.\ \ref{fig:(110)and(100)sus2-300K}(a) and
\ref{fig:(110)and(100)sus2-300K}(b) respectively, but depict the
inverse susceptibility as a function of temperature, which makes
the anomaly at $T_{\rm D} = 120$\,K more visible.  The overall
shape of the susceptibility trace is very similar for ${\bf
H}\parallel$ [110] and ${\bf H}\parallel$ [001].  However, the
anomaly at $T_{\rm D}$ is more pronounced for ${\bf H}\parallel$
[110].  This can be seen in the plots of both $M/H$ vs $T$ and
$(M/H)^{-1}$ vs $T$.

\begin{figure}[!ht]
\begin{center}
\includegraphics{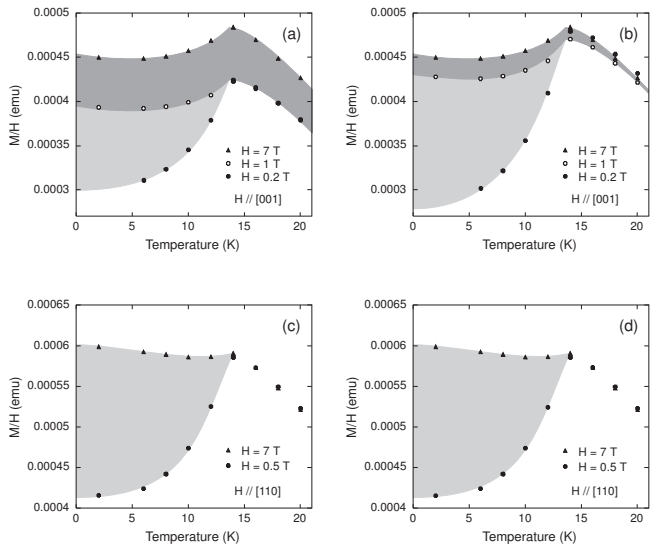}
\caption{Field-induced changes in the temperature-dependence of
the susceptibility.  The field is increased and decreased in
steps, with each new field applied at $T$ = 20\,K. The
susceptibility is measured as a function of $T$ as the crystal is
cooled through $T_{\rm N}$. (a) and (b) show $M/H$ for ${\bf
H}\parallel$ [001]. The field increases in (a) and decreases in
(b). (c) and (d) show $M/H$ for ${\bf H}\parallel$ [110]. The
field increases in (c) and decreases in (d).}
\label{fig:(110)and(100)sus2-20K}
\end{center}
\end{figure}

To investigate the field-dependence of the susceptibility, we made
a series of measurements between 2\,K and 20\,K at different
applied fields.  The field was increased and decreased in steps
from $H$ = 0\,T to $H$ = 7\,T.  For each measurement we applied
the field above $T_{\rm N}$ (at 20\,K), then measured the
susceptibility as a function of temperature while cooling through
$T_{\rm N}$.  The data are shown in Fig.\
\ref{fig:(110)and(100)sus2-20K}, with field increasing in
\ref{fig:(110)and(100)sus2-20K}(a) and
\ref{fig:(110)and(100)sus2-20K}(c) and decreasing in
\ref{fig:(110)and(100)sus2-20K}(b) and
\ref{fig:(110)and(100)sus2-20K}(d).  For both field directions the
susceptibility below $T_{\rm N}$ increases smoothly with field
through the pale grey region (see
\ref{fig:(110)and(100)sus2-20K}(a) and
\ref{fig:(110)and(100)sus2-20K}(c)).  Plots
\ref{fig:(110)and(100)sus2-20K}(c) and
\ref{fig:(110)and(100)sus2-20K}(d) are identical, showing that,
for ${\bf H}\parallel$ [110] the increase below $T_{\rm N}$ is
completely reversible and the susceptibility above $T_{\rm N}$ is
independent of field.  For ${\bf H}\parallel$ [001], a change
occurs at $H = 1$\,T.  Below this field the susceptibility above
$T_{\rm N}$ is independent of field, but above $H = 1$\,T the
whole trace increases irreversibly through the dark grey region
shown in \ref{fig:(110)and(100)sus2-20K}(a).  When the field is
removed, the whole trace decreases through the smaller dark grey
region shown in \ref{fig:(110)and(100)sus2-20K}(b).  As the field
is decreased below $H = 1$\,T the susceptibility above $T_{\rm N}$
becomes independent of field and the trace below $T_{\rm N}$
decreases through the pale grey region.

It should be noted that repetitions of the susceptibility
measurements at low applied fields revealed small, random shifts
whenever the applied field was changed between measurements.  This
was attributed to fluctuations in the magnetisation of the sample
surroundings and trapped flux in the superconducting magnet,
leading to fluctuations in the applied field of 0.001--0.02\,T.
However, at higher fields the fluctuations became negligible
compared to the total field, so the susceptibility was unaffected.
Unfortunately it was impossible to measure the magnetisation of
the sample surroundings, and it was therefore impossible to
measure the exact applied field.  This meant that the magnitude of
the susceptibility at applied fields of $H \leq 0.2$\,T had an
uncertainty of at least $\pm$ 10\%.

\begin{figure}[!ht]
\begin{center}
\includegraphics{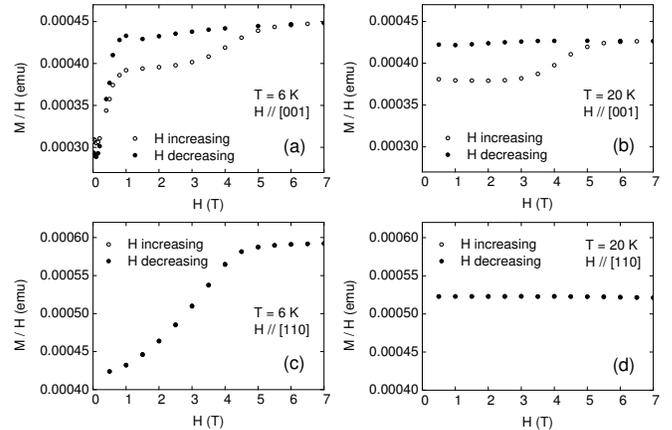}
\caption{Change in the field-dependence of the susceptibility with
temperature. These plots are taken from the same data set as those
in Fig.\ \ref{fig:(110)and(100)sus2-20K}.  Each time the applied
field was changed, the crystal was warmed to $T$ = 20\,K, then
cooled through $T_{\rm N}$ in the new applied field. (a) and (b)
show $M/H$ for ${\bf H}\parallel$ [001] below and above $T_{\rm
N}$ respectively. (c) and (d) show $M/H$ for ${\bf H}\parallel$
[110] below and above $T_{\rm N}$.}
\label{fig:(110)and(100)sus0-7T}
\end{center}
\end{figure}

The plots in Fig.\ \ref{fig:(110)and(100)sus0-7T} are taken from
the same data set as those in Fig.\
\ref{fig:(110)and(100)sus2-20K}, but this time the susceptibility
is plotted as a function of applied field instead of temperature.
Figures \ref{fig:(110)and(100)sus0-7T}(a) and
\ref{fig:(110)and(100)sus0-7T}(c) show the field dependence of the
susceptibility below $T_{\rm N}$, whereas Figs.\
\ref{fig:(110)and(100)sus0-7T}(b) and
\ref{fig:(110)and(100)sus0-7T}(d) show the field dependence above
$T_{\rm N}$.  For ${\bf H}\parallel$ [001] the susceptibility
below $T_{\rm N}$ increases rapidly from $H$ = 0--1\,T, then
increases slowly and irreversibly from $H$ = 1--7\,T (Fig.\
\ref{fig:(110)and(100)sus0-7T}(a)). The susceptibility above
$T_{\rm N}$ changes little between $H$ = 0\,T and $H$ = 1\,T, but
increases irreversibly between $H$ = 3\,T and $H$ = 7\,T (Fig.\
\ref{fig:(110)and(100)sus0-7T}(b)).  For ${\bf H}\parallel$ [110]
the susceptibility below $T_{\rm N}$ increases rapidly and
reversibly between $H$ = 0\,T and $H$ = 7\,T, but is almost
independent of field above $T_{\rm N}$.

\begin{figure}[!ht]
\begin{center}
\includegraphics{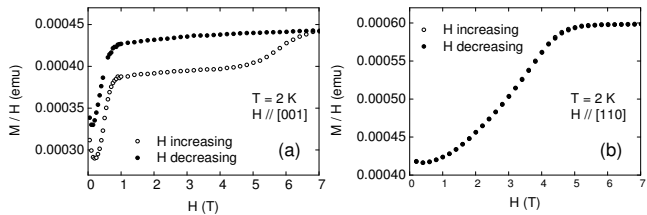}
\caption{Field-dependence of the susceptibility for field applied
at constant $T$ = 2\,K. (a) ${\bf H}\parallel$ [001], (b) ${\bf
H}\parallel$ [110].  The small upturns in the susceptibility at
low fields in both (a) and (b) are probably due to underestimation
of the applied field strength, due to magnetisation of the sample
surroundings.}
\label{fig:(110)and(100)sus0-7T2K}
\end{center}
\end{figure}

In the measurements described above, the field was always applied
above $T_{\rm N}$.  However, we also wished to investigate the
field dependence of the susceptibility below $T_{\rm N}$ without
cooling through $T_{\rm N}$ for each change in field.  For this
measurement the magnetic field was swept while keeping the
temperature constant at 2\,K.  Figure
\ref{fig:(110)and(100)sus0-7T2K} shows the results obtained.  For
${\bf H}\parallel$ [110] the trace is very similar to that
obtained when the crystal was cooled through $T_{\rm N}$ in the
field. However, for ${\bf H}\parallel$ [001] there are two
differences. Firstly, the irreversible increase occurs at a higher
field.  Secondly, on removal of the field at $T$ = 2\,K, the
susceptibility does not return to its original zero field value.
This contrasts with the data shown in Fig.\
\ref{fig:(110)and(100)sus0-7T}(a), where removal of the field
above $T_{\rm N}$ reduced the susceptibility below its original
zero field value.

\begin{figure}[!ht]
\begin{center}
\includegraphics{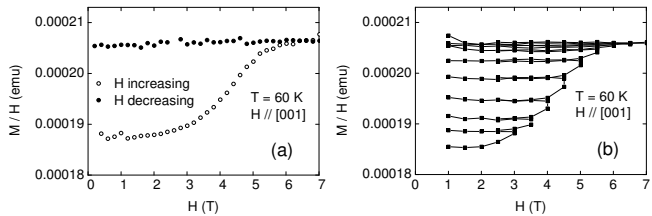}
\caption{Field-dependence of the susceptibility for ${\bf
H}\parallel$ [001], field applied at constant $T$ = 60\,K. The
data in (a) has been corrected for a small remanent field of
$H_{\rm R}$ = 0.008\,T, probably caused by magnetisation of the
sample surroundings or trapped flux in the magnet. (b) shows that
any increase in susceptibility, no matter how small, is
irreversible.}
\label{fig:M(H)60K}
\end{center}
\end{figure}

To see if an irreversible increase in the susceptibility occurred
at higher temperatures we measured the field dependence at $T$ =
60\,K, with ${\bf H}\parallel$ [001]. An irreversible increase did
occur (see Fig.\ \ref{fig:M(H)60K}(a)), but of smaller magnitude
than that observed at $T$ = 20\,K.  To find the field required to
make the increase irreversible, we warmed the crystal above
$T_{\rm D}$ and re-cooled to $T = 60$\,K in zero field.  We then
swept the field up and down several times, reaching successively
larger values each time.  We found that any increase in
susceptibility, no matter how small, was irreversible (see Fig.\
\ref{fig:M(H)60K}(b)).  For ${\bf H}\parallel$ [110] we found the
susceptibility to be completely independent of field at $T$ =
60\,K. The data in Fig.\ \ref{fig:M(H)60K}(a) have been corrected
for a remanent field of $H_{\rm R}$ = 0.008\,T, which was probably
caused by magnetisation of the sample surroundings or trapped flux
in the superconducting magnet. $H_{\rm R}$ was determined by
finding the $y$-intercept of a straight line fitted to a plot of
magnetisation $M$ versus field $H$ ($M = \frac{M}{H}(H + H_{\rm
R})$, where $H_{\rm R}$ is the remanent field).

\begin{figure}[!ht]
\begin{center}
\includegraphics{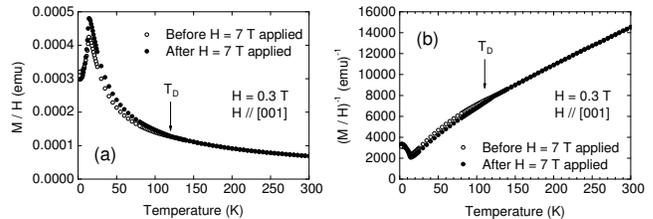}
\caption{Temperature-dependence of $M/H$ at $H = 0.3$\,T for ${\bf
H}\parallel$ [001] before and after application and removal of a
7\,T field. (a) $M/H$ vs $T$. The two traces coincide above
$T_{\rm D}$. (b) $(M/H)^{-1}$ vs $T$.  The anomaly at $T_{\rm D}$
is more pronounced after application and removal of the 7\,T
field.}
\label{fig:(100)M(T)BeforeAfter}
\end{center}
\end{figure}

The irreversible increase in susceptibility for ${\bf H}\parallel$
[001] persists up to $T_{\rm D}$. This is shown in Fig.\
\ref{fig:(100)M(T)BeforeAfter}(a), where the temperature
dependence of the susceptibility, measured at $H$ = 0.3\,T is
plotted before and after application and removal of a 7\,T field.
The two traces coincide for $T > T_{\rm D}$. Figure
\ref{fig:(100)M(T)BeforeAfter}(b) shows the inverse susceptibility
(taken from the same data set as Fig.\
\ref{fig:(100)M(T)BeforeAfter}(a)).  The anomaly at $T_{\rm D}$ is
more pronounced after application and removal of the 7\,T field.

\subsection{Summary}

The main results of the susceptibility studies are as follows. The
anomaly at $T_{\rm D}$ is more pronounced when the field is
applied along [110] than when it is applied along [001]. For ${\bf
H}\parallel$ [110] the susceptibility increases reversibly with
field below $T_{\rm N}$, but is independent of field above $T_{\rm
N}$. For ${\bf H}\parallel$ [001] the susceptibility increases
rapidly below $T_{\rm N}$ from $H$ = 0\,T to $H$ = 1\,T and more
gradually from $H$ = 1\,T to $H$ = 7\,T. The increase from 0--1\,T
is quasi-reversible and occurs only below $T_{\rm N}$, whereas the
increase from 1--7\,T is irreversible, and also occurs above
$T_{\rm N}$. The field strength at which the latter occurs depends
on whether the field is applied below or above $T_{\rm N}$.  For
fields applied at constant $T < T_{\rm N}$ a field of 7\,T is
required to achieve the full increase in susceptibility, whereas
for fields applied above $T_{\rm N}$, only 5.5\,T is required. Any
increase in susceptibility above $T_{\rm N}$, no matter how small,
is irreversible.  Above $T_{\rm D}$ the susceptibility is
independent of field.

\section{Interpretation of susceptibility data}

We now attempt to interpret the susceptibility data using our
hypothesis that the formation of magnetic domains is strongly
influenced by the distribution of structural domains.

The most likely cause of the reversible increase in susceptibility
with field for ${\bf H} \parallel [110]$ is the proposed
reversible spin reorientation in the type-I component of the
magnetic structure.

For ${\bf H} \parallel [001]$ we propose that the rapid
quasi-reversible increase in susceptibility between $H = 0$\,T and
$H = 1$\,T is due to initial favouring of magnetic domains whose
spins are perpendicular to the field ($D_a^b$, $D_b^a$, $D_c^a$
and $D_c^b$).  This would give rise to a small decrease in the
magnetic structure factor of the (100) Bragg peak and a small
increase in the magnetic structure factor of the (110) Bragg peak,
in reasonable agreement with the small changes in Bragg
intensities observed at $H = 1$\,T (see Fig.\
\ref{fig:D10(010)and(110)vsH}). Between $H = 1$\,T and $H = 7$\,T
we propose that the magnetic field becomes large enough to
influence the alignment of the structural domains.  The resulting
structural domains cause those type-I magnetic domains whose
ordering vectors are perpendicular to ${\bf H}$ to be favoured.
This leaves only domains $D_a^b$ and $D_b^a$, causing the magnetic
structure factors, and hence the intensities of the Bragg
reflections, to change.  When the field is decreased below 1\,T,
we believe that domains $D_a^c$ and $D_b^c$ return. Combining the
structure factors of these domains with those of $D_a^b$ and
$D_b^a$ makes no difference to the average structure factors, in
agreement with the observed absence of change in Bragg intensities
on removal of the field.  However, the return of these domains
with spins parallel to the applied field would be expected to
cause a decrease in susceptibility, as observed.

The above discussion shows that the hypotheses outlined in Section
\ref{sec:domains} provide some success in explaining the
susceptibility data.  However, some observations remain
unexplained.  For instance, we observe an irreversible increase in
susceptibility between $H = 1$\,T and $H = 7$\,T for ${\bf H}
\parallel [001]$ both below and above $T_{\rm N}$.  We assume that
the increase above $T_{\rm N}$ is due to the effect of the applied
field on the ordering of the Pr orbitals.  However, no
irreversible increase is observed either above or below $T_{\rm
N}$ for ${\bf H} \parallel [110]$, and we do not understand why
this should be. We also do not understand why a change in domain
populations between $H = 0$\,T and $H = 1$\,T should cause a
change in susceptibility for ${\bf H} \parallel [001]$ when the
irreversible change in domain populations after application of a
large field along [110] causes no irreversible change in the
susceptibility below $T_{\rm N}$.

Despite the above problems, the interpretation of the
susceptibility data is in broad agreement with that of the neutron
diffraction data.  We also note that the susceptibility
measurements at low fields indicate that the [001] direction is
the easy direction of magnetisation, in support of a single-${\bf
q}$ type-I component of the magnetic structure.

\section{Conclusion}

We have presented (i) neutron diffraction studies of the
antiferromagnetic structure of single crystal PrO$_2$ as a
function of applied field (ii) susceptibility measurements on
single crystal PrO$_2$ as a function of applied field and
temperature. Both studies were carried out for fields applied
along two different symmetry directions: [001] and [110].

The neutron diffraction studies revealed irreversible changes in
the magnetic Bragg intensities after application and removal of a
large field.  We attributed these changes to magnetic domain
alignment, caused by a field-induced change in the underlying
distribution of structural domains. For ${\bf H}
\parallel [0\bar{1}1]$ we also observed certain reflections to
disappear at high field and reappear when the field was removed.
We attributed this observation to a spin reorientation of the
type-I component of the magnetic structure.

The susceptibility studies revealed a reversible increase with
applied field below $T_{\rm N}$ along both symmetry directions, as
well as an irreversible increase at higher fields both above and
below $T_{\rm N}$ for ${\bf H} \parallel [001]$.  The irreversible
increase disappeared at $T_{\rm D}$, the temperature at which a
cooperative Jahn-Teller distortion of the oxygen sublattice
occurs.  The changes in susceptibility with applied field are in
broad agreement with the proposed alignment of structural and
magnetic domains, although some features remain poorly understood.

We conclude that our neutron diffraction and susceptibility
studies have revealed striking hysteresis effects on the
application and removal of a large magnetic field below $T_{\rm
D}$. We propose that the field interacts strongly with the lattice
to influence the alignment of structural domains below $T_{\rm
D}$, and that this in turn influences the alignment of the
magnetic domains below $T_{\rm N}$.

\begin{acknowledgments}
We would like to thank P. Santini for insightful discussions and
F. Wondre for help with sample preparation. Financial support and
provision of a studentship for CHG by the EPSRC is also
acknowledged.
\end{acknowledgments}

\bibliography{PrO2PrBCO}

\end{document}